\documentclass[lineno]{jfm}
\usepackage{graphicx}
\usepackage{epstopdf, epsfig}

\usepackage{newtxtext}
\usepackage{newtxmath}
\usepackage{natbib}
\usepackage{hyperref}
\usepackage{xcolor}

\hypersetup{
    colorlinks = true,
    urlcolor   = blue,
    citecolor  = black,
}
\newcommand{\RomanNumeralCaps}[1]

\shorttitle{Particle manipulation in internal Stokes flow}
\shortauthor{X. Liu,
  PK. Das
 and S. Hilgenfeldt }

\title{Principles of hydrodynamic particle manipulation in internal Stokes flow}

\author{Xuchen Liu,
  Partha Kumar Das
 \and Sascha Hilgenfeldt \corresp{\email{sascha@illinois.edu}}}

\affiliation{Department of Mechanical Science and Engineering, The Grainger College of Engineering, University of Illinois Urbana-Champaign,
Urbana, Illinois 61801, USA}

\begin{document}

\maketitle

\begin{abstract}
Manipulation of small-scale particles across streamlines is the elementary task of microfluidic devices. Many such devices operate at very low Reynolds numbers and deflect particles using arrays of obstacles, but a systematic quantification of relevant hydrodynamic effects has been lacking. Here, we explore an alternate approach, rigorously modeling the displacement of force-free spherical particles in vortical Stokes flows under hydrodynamic particle-wall interaction. Certain Moffatt-like eddy geometries with broken symmetry allow for systematic deflection of particles across streamlines, leading to particle accumulation at either Faxen field fixed points or limit cycles. Moreover, particles can be forced onto trajectories approaching channel walls exponentially closely, making quantitative predictions of particle capture (sticking) by short-range forces possible. This rich, particle size-dependent behavior suggests the versatile use of inertial-less flow in devices with a long particle residence time for concentration, sorting, or filtering.  

\end{abstract}

\begin{keywords}
Authors should not enter keywords on the manuscript, as these must be chosen by the author during the online submission process and will then be added during the typesetting process (see http://journals.cambridge.org/data/\linebreak[3]relatedlink/jfm-\linebreak[3]keywords.pdf for the full list)
\end{keywords}

\section{\label{sec:level1}Introduction}

Controlled manipulation of small particles in suspension is crucial in fundamental
research and applications such as biomedical and biochemical
processing \citep{ateya2008good, nilsson2009review}, disease diagnostics and therapeutics \citep{gossett2010label, puri2014particle}, drug discovery and delivery systems \citep{dittrich2006lab, nguyen2013design}, self-cleaning and antifouling technologies \citep{callow2011trends, kirschner2012bio}. The essence of particle manipulation is to drive the particles across streamlines, making them follow specific pathlines (trajectories) distinct from the fluid elements based on their properties. 

Microfluidic particle manipulation aims at transportation, separation, trapping, and enrichment \citep{sajeesh2014particle,lu2017particle} and can be achieved through various approaches. Many techniques exploit certain particles' response to external forces, e.g.\ electrical \citep{xuan2019recent}, optical \citep{lenshof2010continuous}, and magnetic techniques \citep{van2014integrated}. However, not all particles of interest will be susceptible to these, which is why there is a continued interest in manipulation based solely on hydrodynamic forces \citep{karimi2013hydrodynamic}. Most notably, techniques that use particle inertia have gained prominence \citep{di2007continuous,di2009inertial,agarwal2018inertial} and quantitative theories have been developed beyond classical equations of motion \citep{maxey1983equation} to rigorously describe the effect of inertial forces in both the background flow and the disturbance flow around the particle \citep{agarwal2021unrecognized,agarwal2021rectified}. \textcolor{black}{Recent work by  \citet{agarwal2024density} also integrates the important case of acoustofluidic particle manipulation \citep{bruus2011forthcoming, laurell2007chip, friend2011microscale} as a particular limit of inertial particle manipulation.} Despite the description of such forces in simple flow fields that is now known analytically, many practical cases still lack a fundamental quantitative theory on how devices based on hydrodynamic effects work.

This also applies to viscous Stokes flow. Even in the absence of inertia, particles interact hydrodynamically with other particles or large-scale interfaces (walls or fluid-fluid boundaries), with effective interactions that are notoriously long-ranged \citep{happel1965low, brady1988stokesian, kim2013microhydrodynamics, pozrikidis1992boundary, pozrikidis2011introduction}. Early theoretical efforts by \citet{brady1988stokesian,claeys1989lubrication,claeys1993suspensions} show in general terms that a particle moving in a Stokes flow should never experience surface-to-surface contact with a boundary (interfaces cannot touch in finite time). However, in practical situations where Stokes flow around obstacles is used to manipulate particles such as DLD (deterministic Lateral Displacement) \citep{huang2004continuous, mcgrath2014deterministic, zhang2020concise}, most modeling descriptions assume contact with obstacles and eschew any proper hydrodynamic modeling. Very recent, more careful studies of the interaction between non-spherical particles with obstacles in Stokes flow \citep{li2024dynamics} describe trajectories without contact while still observing a net displacement effect on the transported particle. However, that work uses an ad hoc interaction force \citep{dance2004collision} rather than the full hydrodynamic interaction between the particle and the interface. 

In all cases, a single encounter of a particle with an obstacle has a very small net effect on particle position \citep{li2024dynamics, partha2024control}, which is why practical DLD set-ups use forests of pillar obstacles. Therefore, this work focuses on vortical flows that enable repeated particle-interface encounters for sizable cumulative effects. In Sec.~\ref{sec: interaction}, we present modeling equations for particle displacement by hydrodynamic interactions in Stokes flow. In Sec.~\ref{sec: quantify results}, we quantify the results in analytically known internal Stokes flows, suggesting novel design strategies for precisely manipulating particles in Stokes flow, from accumulation to capture. Section~\ref{sec: conclusion} provides discussion and conclusions.

\section{HYDRODYNAMIC MODEL OF PARTICLE-WALL INTERACTION IN
STOKES FLOW}\label{sec: interaction}
\begin{figure}
    \centering
\includegraphics[height=5.5cm]{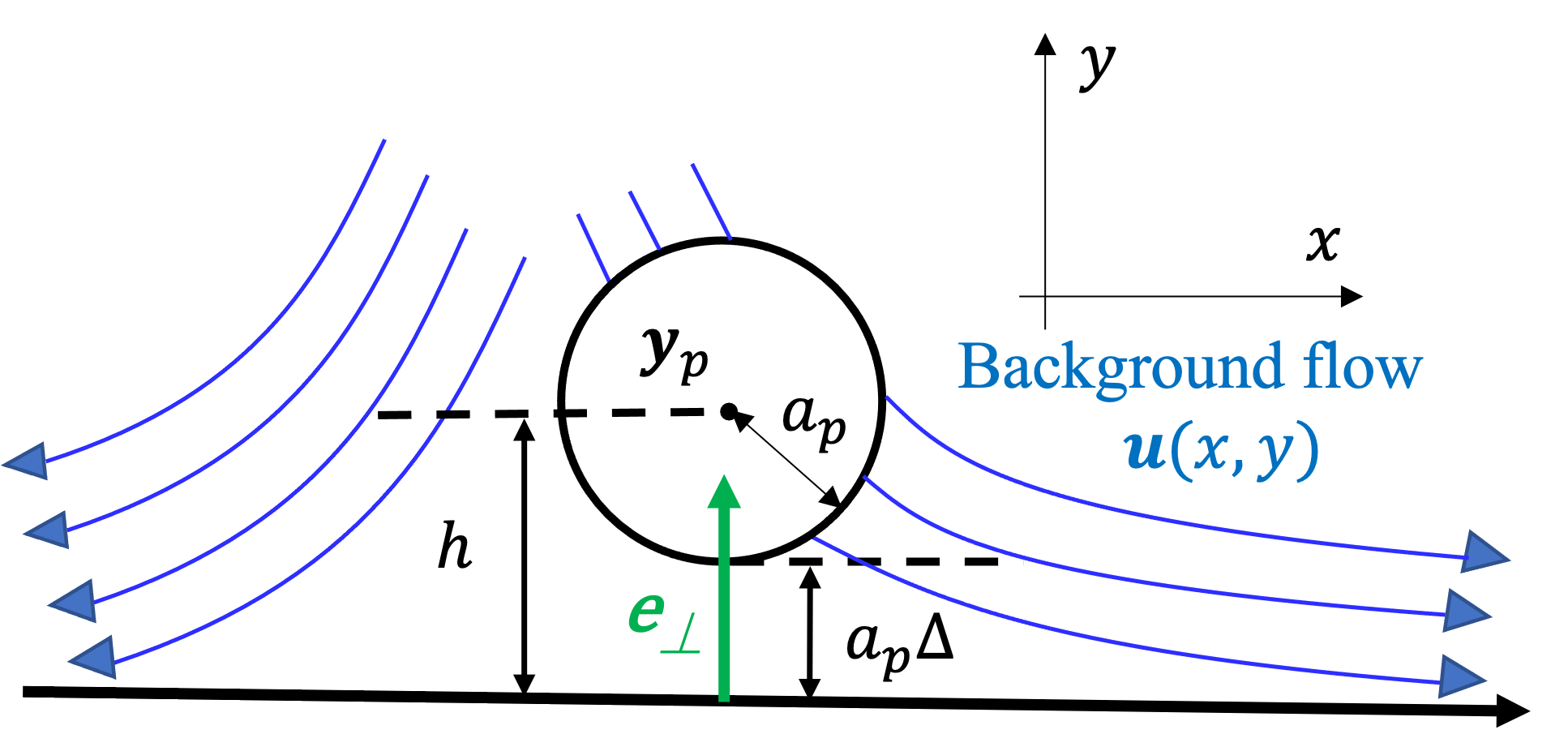}
    \caption{Schematic of a particle at $(x_p,y_p)$ near a flat wall submerged in an arbitrary background flow ${\bf u}$.}
    \label{Fig: Parallel force shematic}
\end{figure}
In most microfluidic set-ups, at least a subset of particles placed into the flow will be located or transported near boundaries, which could be solid (no-slip) or fluid/fluid interfaces (e.g.\ immiscible liquids, droplets, bubbles). Generally, the proximity of a boundary will lead to specific displacements depending on Reynolds number, density contrast, and other parameters. In steady inertial microfluidics the presence of boundaries and the resulting flow gradients lead to slow inertial migration even at considerable distances to the boundaries \citep{segre1962behaviour, di2009inertial}, while in oscillatory inertial microfluidics (such as set-ups using acoustically driven microbubbles) boundary effects become important in very close proximity and can be approximately treated by lubrication theory \citep{thameem2017fast,agarwal2018inertial}. In flow with negligible inertia, one would expect boundary effects to be longer-range and potentially more prominent, but the question of whether a practically usable net displacement after an encounter of a particle with a wall or an obstacle is feasible has not been fundamentally answered. 

Different quantities can be targeted in modeling of particles in Stokes flow, particularly (i) the forces on a particle moving at a given speed, (ii) the forces on a particle held fixed in a certain location, or (iii) the motion of a force-free particle. The latter is our focus here, as it describes the trajectory of a density-matched particle not subject to external forces. Progress in describing all three cases has built on a body of literature based on early pioneering work \citep{brenner1961slow,goldman1967slow,goldman1967slow2}. 

Generally, studies of force-free particles result in predictions for the deviation of the particle velocity from the background fluid velocity it is embedded in, i.e., the non-passive part of the particle motion. This velocity correction can be decomposed into effects in the direction parallel to the boundary and in the perpendicular direction. 
A comprehensive analysis of the equation of motion of a force-free particle with wall-normal velocity corrections is given by \citet{rallabandi2017hydrodynamic}. Other work has described wall-parallel velocity corrections far from and near the wall \citep{ekiel2006accuracy, pasol2011motion}. 
\textcolor{black}{ 
Here, we derive an original formalism for the displacement
of spherical particles in Stokes flows, expanding on existing work to arrive at an equation of motion applicable for all particle-wall distances. We will demonstrate several modes of systematic (net) particle displacement across streamlines due to wall interaction effects, a phenomenon not previously acknowledged widely. While particles can never cross streamlines in unidirectional Stokes flow \citep{bretherton1962motion}, streamline crossing should generally be expected in the presence of wall-normal flow components. A simple example is a particle very close to a wall transported in a channel flow undergoing contraction -- the particle cannot stay on its initial streamline without penetrating the wall.}


\subsection{Moffatt Eddies}

\begin{figure}
    \centering
\includegraphics[height=4cm]{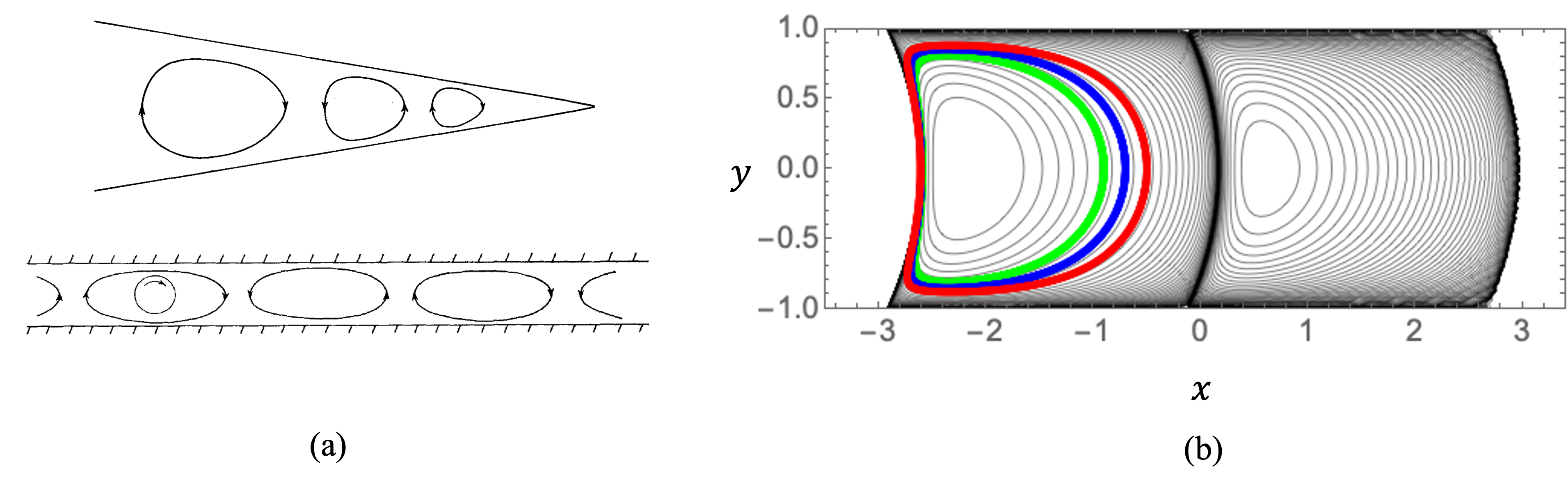}
    \caption{(a) Schematic of Stokes flow in a wedge between rigid boundaries (top) and between parallel plates (bottom); the source of the fluid motion is a rotating cylinder between the planes. Modified from \citep{moffatt1964viscous}. (b) The eddy streamline pattern in the latter case, from the symmetric stream function Eq.~\eqref{streamfunction symmetric moffatt}. Particles ($a_p=0.1$) follow closed trajectories (colored) for different particle initial positions.}
    \label{Fig: Symmetric Moffatt}
\end{figure}

The ideal test case to quantify boundary effects in a Stokes flow is a flow that is (i) analytically known and for which (ii) the walls are isolated and flat. We take inspiration from the classic work of \citet{moffatt1964viscous}: when two rigid flat boundaries form a wedge, a distant stirring of the fluid will induce a flow consisting of a sequence of vortices shown in the upper panel of Figure \ref{Fig: Symmetric Moffatt} (a). 
\citet{moffatt1964viscous} also describes the special case of zero wedge angle, i.e., a Stokes flow between two parallel plates as sketched in the lower panel of Figure \ref{Fig: Symmetric Moffatt} (a). We set these two plates at $y=\pm 1$.

The \textcolor{black}{Moffatt} parallel plate solution consists of a series of alternating congruent vortices that take up the height of the channel (Fig.~\ref{Fig: Symmetric Moffatt}(b) shows streamline contours) and whose strength decays exponentially with distance from a stirrer on the far left. The corresponding stream function $\psi$ is known analytically (asymptotically far from the stirrer) and has the form
\begin{equation}
    \psi = (A \cos ky+By \sin ky)e^{-kx}\,.
    \label{streamfunction symmetric moffatt}
\end{equation}
The constant $A$ is an overall scale, which can be set to one. In order to fulfill the no-slip boundary conditions at $y=\pm 1$, $B=-\cot(k)$ follows. Additionally, the complex parameter $k=p+iq$ must satisfy the transcendental equation $2k+\sin 2k=0$ \citep{moffatt1964viscous}. The solution with the smallest positive real part is $p\approx 2.106, q\approx 1.125$. Its symmetry with respect to the center plane of the channel implies (together with the time-reversal symmetry of Stokes flow) that a particle released anywhere and deflected by interaction with one of the walls will experience the opposite deflection when encountering the other wall so that all trajectories must close and there is no net displacement of particles. Figure~\ref{Fig: Symmetric Moffatt} (b) shows some examples computed with the deflection formulae derived in Sec.~\ref{subsection: wall parallel particle} and \ref{subsection: wall normal particle}, but the statement is true independent of the exact formalism.
Thus, we will focus instead on a different analytical solution of the parallel-plate Moffatt case whose stream function is antisymmetric in $y$,
\begin{equation}
    \psi = (C \sin ky+ Dy \cos ky)e^{-kx}
    \label{streamfunction antisymmetric moffatt}
\end{equation}
Again, we set $C=1$ and $D=-\tan(k)$ to fulfill boundary conditions at $y=\pm 1$. 
The parameter $k$ must now satisfy the transcendental equation
\begin{equation}
    2k-\sin 2k=0
\end{equation}
and the relevant solution with the smallest positive real part is $k=p+iq$ with $p=3.749$, $q=1.384$. This flow, with two vortices across the channel, is shown in Fig.~\ref{Fig3}(a). The vortex-to-vortex distance in the $x$ direction is thus $\xi=\pi/q\approx 2.27$, and the damping factor of the flow speeds in neighboring vortices is $\zeta=e^{p\pi/q}=e^{8.51}\approx4950$. 
We will see that this flow accomplishes permanent net displacements of particles. 

\subsection{Particle velocity in the presence of a wall}
In any ambient Stokes flow, the velocity of a spherical particle of radius $a_p$ differs from the background flow velocity $\bf{u}$ (without walls) by the Faxen correction \citep{faxen1922widerstand} evaluated at the particle position $\bf{x}_p$, resulting in
\begin{equation}
   \boldsymbol{u}_{p,F}(\boldsymbol{x}_p(t))  = \boldsymbol{u}(\boldsymbol{x}_p(t)) + \frac{a_p^2}{6}\nabla^2 \boldsymbol{u}(\boldsymbol{x}_p(t))
   \label{Faxen velocity}
\end{equation}

For finite distances $h$ between the center of the particle and the wall (cf. Fig.~\ref{Fig: Parallel force shematic}), this velocity is modified by the presence of the wall. Both particle and fluid inertia are absent, and the particle trajectory is described by a first-order overdamped dynamical system with the wall interaction effects as a velocity correction $\boldsymbol{W}(h)$,
\begin{equation}
    \frac{d\boldsymbol{x}_{p}(t)}{dt} = \boldsymbol{u}_p(t)= \boldsymbol{u}_{p,F}(\boldsymbol{x}_p(t))+\boldsymbol{W}(\boldsymbol{x}_p(t),h)\,.
    \label{force free motion}
\end{equation}

We will show that in many situations for small $a_p$ the wall effect $\boldsymbol{W}$ is perturbative, i.e., of a higher order than $a_p^2$. The principle of (\ref{force free motion}) has been acknowledged in the literature \citep{brenner1961slow, 
 goldman1967slow, goldman1967slow2, o1964slow, o1967slow, o1967slow2, perkins1992hydrodynamic}, but has not been systematically applied for arbitrary $h$ to determine particle trajectories meant for net displacement.
The following subsections quantify the particle velocity corrections $\boldsymbol{W}$ parallel to and normal to the walls.

\subsection{Wall-parallel corrections to the particle velocity}\label{subsection: wall parallel particle}
Consider first a force-free sphere embedded in a semi-infinite region bounded by a plane no-slip wall at $y=-1$ (cf.\ Fig.~\ref{Fig: Parallel force shematic}), so that $h=y+1$. Decomposing the ambient velocity field $\boldsymbol{u}=(u,v)$, we now focus on corrections $W_x$ to the wall-parallel motion $u_p$. This component of the wall effect is conveniently expressed as a fraction of the Faxen-corrected velocity, i.e., 
\begin{equation}
 W_{x}(x,y)=-f(\Delta) \left(u(x,y)+\frac{a_p^2}{6}\nabla^2 u(x,y)\right)
\label{wall parallel particle motion}
\end{equation}
where we have replaced $h$ by the dimensionless gap measure
\begin{equation}
    \Delta\equiv\frac{h-a_p}{a_p}\,,
    \label{deltadef}
\end{equation}
representing the surface-to-surface distance relative to the radius of the particle, cf.\ \citep{rallabandi2017hydrodynamic, thameem2017fast, agarwal2018inertial}.

The wall-parallel velocity correction coefficient $f(\Delta)$ has been worked out in detail for specific cases such as linear shear flow \citep{o1968sphere, jeffrey1984calculation,  stephen1992characterization, williams1994particle, chaoui2003creeping}, quadratic flow \citep{goren1971hydrodynamic, ekiel2006accuracy, pasol2006sphere}, or modulated shear flow \citep{pasol2006sphere}, with asymptotic expressions available for $\Delta\to 0$ and $\Delta\to\infty$. We note that (i) the wall effects are most prominent for small $\Delta$ and (ii) the linear shear part of any flow dominates as $\Delta\to 0$. In particular, 
when $\Delta=0$ (particle touching the wall), the sphere has to come to rest.


In order to obtain a uniformly valid expression for $f(\Delta)$, we follow the expansion approach of \citep{pasol2011motion} but modify it to enforce exact matching with known asymptotic results. For $\Delta\ll 1$, linear shear flow is dominant, and Williams's near-wall expression \citep{williams1994particle} must be recovered. Far from the wall, $f(\Delta)\to c \Delta^{-3}$, where the positive constant $c={\cal O}(1)$ depends on the type of flow \citep{goldman1967slow2,  ghalia2016sphere}. The exact value of $c$ makes no qualitative difference to the effects explored here, and we enforce $c=5/16$ to agree with the far-field asymptote provided by \citet{goldman1967slow2} for linear shear flow. 

\begin{figure}
    \centering
\includegraphics[height=8.5cm]{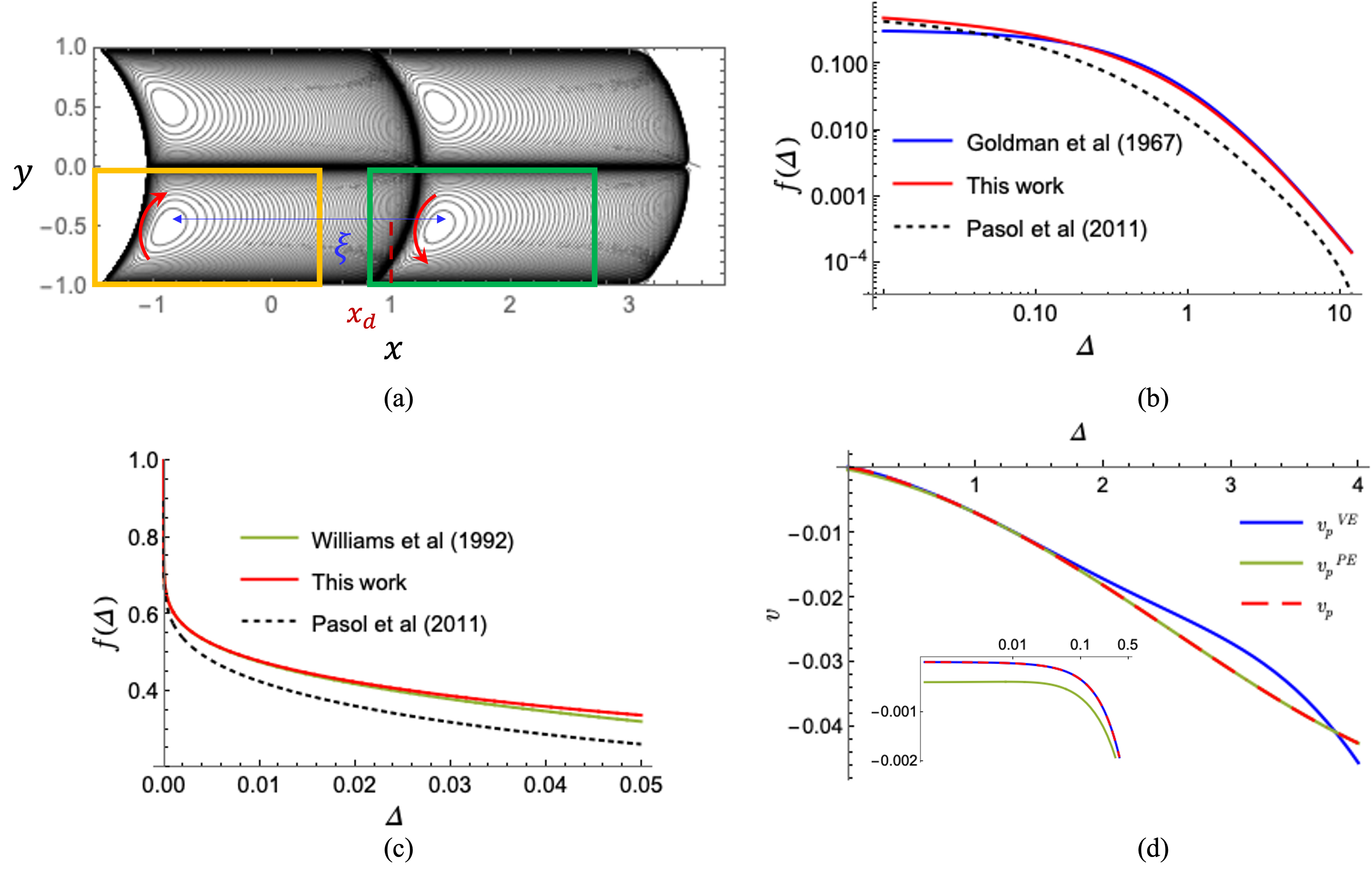}
    \caption{(a) Moffatt eddy flow with pairs of counterrotating vortices taking up the channel height, from the antisymmetric stream function \eqref{streamfunction antisymmetric moffatt}. (b) Wall-parallel flow modification factor $f(\Delta)$ at large $\Delta$. (c) Plot of $f(\Delta)$ at small $\Delta$. (d) Example of normal particle velocity as a function of $\Delta$, for $x=1$ and $a_p=0.1$ \textcolor{black}{in the flow of (a)}: far from the wall, the model follows the particle-expansion velocity $v_p^{PE}$ from \eqref{wallnormalvp}, while for $\Delta<0.5$ (inset) the variable-expansion approach of \eqref{quadexpand} is used.}
\label{Fig3}
\end{figure}

Appendix~\ref{appendix A} details the derivation leading to the following expression:
\begin{equation}
    \label{Us}
f(\Delta)=1-\frac{(1+\Delta)^{4}}{0.66+\Delta(3.15+\Delta(5.06+\Delta(3.73+\Delta)))-0.27(1+\Delta)^{4}\log(\Delta/(1+\Delta))}\,,
\end{equation}
employed for all $\Delta$. Figure~\ref{Fig3}(b,c) illustrate the agreement with the asymptote at $\Delta\gg 1$ \citep{goldman1967slow2} as well as the logarithmic lubrication-theory approach to $f=1$ at $\Delta\to 0$ \citep{stephen1992characterization, williams1994particle}.

Note that this logarithmic behavior means that $1-f$ only drops to $\approx 0.32$ at $\Delta= 10^{-4}$, which for a typical particle of $a_p=5 \mu$m translates into a sub-nanometer gap, where continuum theory breaks down. Thus, in practical situations, $f$ will slow the wall-parallel motion significantly but never dramatically. Furthermore, by its nature, this wall-parallel velocity modification is much less important than the wall-normal effect in pushing particles across streamlines, which is the main focus of the present work. 

\subsection{Wall-normal corrections to the particle velocity}\label{subsection: wall normal particle}
A general expression for the wall-normal component of the hydrodynamic force on spherical particles in arbitrary Stokes flows was obtained by \citet{rallabandi2017hydrodynamic}, 
employing a quadratic expansion of the background flow around the center of the particle: 
\begin{equation}
    \boldsymbol{F_{\perp}}=6\pi\mu a_p[\{-\mathcal{A}(\boldsymbol{u}_p-\boldsymbol{u})-a_p\mathcal{B}(\boldsymbol{e_{\perp}}\cdot \nabla\boldsymbol{u})+\frac{a_p^2}{2}\mathcal{C}(\boldsymbol{e_{\perp}}\boldsymbol{e_{\perp}}:\nabla\nabla\boldsymbol{u})+\frac{a_p^2}{2}\mathcal{D}\nabla^2\boldsymbol{u}\}\cdot\boldsymbol{e_{\perp}}]_{\boldsymbol{x}_p}\,,
    \label{normal force}
\end{equation}
where $\boldsymbol{e_{\perp}}$ is the unit normal to the wall pointing toward the particle center, and $\mu$ is the viscosity of the fluid. $\boldsymbol{u}_p = (u_p, v_p)$ denotes the velocity of the particle when employing an expansion around the particle position.
The scalar quantities $\mathcal{A}$, $\mathcal{B}$, $\mathcal{C}$ and $\mathcal{D}$ are analytically known dimensionless hydrodynamic resistances depending on $\Delta$. The first correction term, proportional to $\mathcal{A}$, is due to the translation of the particle relative to the mean surrounding background flow and is identical to the general expression from \citet{brenner1961slow}. The term proportional to $\mathcal{B}$ is due to extensional gradients of the flow field. This contribution is zero for a sphere in an infinite flow field \citep{happel2012low, batchelor1970stress} but is generally non-zero for a finite distance to the wall. The second moments of the background flow result in two separate contributions to the force: one (proportional to $\mathcal{C}$) dependent on the curvature of the background flow velocity normal to the wall ($\boldsymbol{e_{\perp}}\boldsymbol{e_{\perp}}:\nabla\nabla\boldsymbol{u}$); another (proportional to $\mathcal{D}$) proportional to $\nabla^2\boldsymbol{u}$. This latter term asymptotes to the Faxen correction as $\Delta\to\infty$.

The full analytical expressions for $\mathcal{A}$, $\mathcal{B}$, $\mathcal{C}$ and $\mathcal{D}$ are given in \citet{rallabandi2017hydrodynamic}. The asymptotic behaviors of $\mathcal{A},\mathcal{B},\mathcal{C},\mathcal{D}$ for large separations $\Delta \gg 1$ ($\mathcal{A}_{large}$ etc.)
and for small separations $\Delta \ll 1$ ($\mathcal{A}_{small}$ etc.)
are given in Appendix \ref{appendix B}.

For our case of a force-free particle, we set $\boldsymbol{F_{\perp}}=0$ in \eqref{force free motion} as well as $\boldsymbol{e}_{\perp}=\pm\boldsymbol{e}_y$ (for the wall at $y=\mp 1$, repectively). The resulting equation can be solved for the wall-normal particle velocity $v_p^{PE}$
\textcolor{black}{ -- the superscript PE stands for particle expansion as we expand the background flow around ${\bf x}_p(t)$.
Writing the wall-normal velocity corrections $W_y^\pm$ due to the presence of both walls at $y=\pm 1$ separately, we have}
\begin{equation}
    v_{p}^{PE}({\bf x}_p(t))=v_{p,F}({\bf x}_p(t))+W_y^-({\bf x}_p(t))+W_y^+({\bf x}_p(t))\,,
    \label{wallnormalvp}
\end{equation}
\textcolor{black}{with}
\begin{equation}
    W_y^\pm({\bf x}_p(t))=\pm a_p\frac{\mathcal{B}}{\mathcal{A}}\left. \frac{\partial v}{\partial y}\right|_{\bf{x}_p}+a_p^2\frac{\mathcal{C}}{2\mathcal{A}}\left. \frac{\partial^2 v}{\partial y^2}\right|_{\bf{x}_p}+a_p^2\left(\frac{\mathcal{D}}{2\mathcal{A}}-\frac{1}{6}\right) \left. \nabla^2 v\right|_{\bf{x}_p} \,,
    \label{wyfull}
\end{equation}
and it is understood that $\mathcal{A}, \mathcal{B}, \mathcal{C}, \mathcal{D}$ are evaluated at arguments $\Delta$ from \eqref{deltadef} defined by $h=1\pm y_p$ for the walls at $y=\mp 1$, respectively.
Note that the last term of \eqref{wyfull} explicitly subtracts the Faxen correction so that each $W_y$ term vanishes as $\Delta\to\infty$. 
When the particle approaches a wall closely, we enter the regime of $\Delta\ll 1$. 
In the wall-parallel direction, we still use the expression for $W_x$ from equation \eqref{wall parallel particle motion}, which includes the lubrication limit at very small $\Delta$. However, in the wall-normal direction, the formalism relying on Taylor expansion of the flow around the particle center $v_{p}^{PE}$ is not accurate enough to describe the particle motion -- this is easily seen because the predicted particle normal velocity from \eqref{wallnormalvp} does not vanish when the particle touches the wall. In \citep{rallabandi2017hydrodynamic}, it was shown that for $\Delta\ll 1$, the no-penetration boundary condition can be enforced by replacing the background flow field in \eqref{wyfull} by its quadratic expansion around the point on the wall closest to the particle. 
However, we find that this wall-expansion formalism will not smoothly transition to the expression \eqref{wyfull} when $\Delta\sim 1$, \textcolor{black}{because the exponential dependence of the flow field \eqref{streamfunction antisymmetric moffatt} with substantial $|k|$ compromises the accuracy of the quadratic expansion even at relatively short distances from the wall.} 

\textcolor{black}{Therefore, we generalize and improve the transition to $v_{p}^{PE}$ at small $\Delta$ by constructing the quantity $v^{VE}(x,y)$, the second-order expansion of $v(x,y)$ around variable expansion points} $x_E=x_p, y_E=y_E(y_p)$, i.e., 
\begin{equation}
    v^{VE}(x,y)=v(x_E,y_E)+ (y-y_E) \left.\frac{\partial v}{\partial y}\right\vert_{{\bf x}_E}+\frac{1}{2}(y-y_E)^2\left. \frac{\partial^2 v}{\partial y^2}\right\vert_{{\bf x}_E}+\frac{1}{2}(x-x_E)^2\left. \frac{\partial^2 v}{\partial x^2}\right\vert_{{\bf x}_E}\,.
    \label{quadexpand}
\end{equation}
We omit the linear in $x$ and mixed terms because these  never give nonzero contributions when evaluated within our formalism.
The expansion point must coincide with the nearest point on the wall when the particle is touching, and with the particle $y$-position as $\Delta\to 1$ to consistently merge into the particle-expansion formalism. Thus, for a wall at $y=-1$ and a particle at $y_p$ we set
\begin{equation}
y_E = 1+2(y_p-a_p)\,.
    \label{yeofy}
\end{equation}
For $\Delta<1$, we use this variable expansion point velocity $v^{VE}$ instead of $v$ in the evaluation of particle velocity normal to the wall, resulting in 
\begin{equation}
    v_{p}^{VE}({\bf x}_p(t))=v^{VE}({\bf x}_p)+\frac{a_p^2}{6} \nabla^2 v^{VE}({\bf x}_p)+W_y^{VE-}({\bf x}_p)\,.
    \label{wallnormalvpmodify}
\end{equation}
Note that the effects of the wall at $y=+1$ are negligible here.
The wall correction is now 
\begin{equation}
    W_y^{VE-}({\bf x}_p)= -a_p\frac{\mathcal{B}}{\mathcal{A}} \left.\frac{\partial v^{VE}}{\partial y}\right|_{\bf{x}_p}+a_p^2\frac{\mathcal{C}}{2\mathcal{A}}\left. \frac{\partial^2 v^{VE}}{\partial y^2}\right|_{\bf{x}_p}+a_p^2\left(\frac{\mathcal{D}}{2\mathcal{A}}-\frac{1}{6}\right)\left.\nabla^2 v^{VE}\right|_{\bf{x}_p}\,,
    \label{wyVE}
\end{equation}
where the derivatives and resistance coefficients are still evaluated at the particle position.
This formalism smoothly interpolates between no-penetration for $\Delta\ll 1$ and the second-order approximation to $v^{PE}$ at $\Delta=1$. The eventual particle velocity normal to the wall for any $\Delta$ is taken to be piecewise,
\begin{equation}
v_p(x,y) =
\begin{cases}
v_p^{PE} \qquad \text{if} \quad\Delta \geq 1 \\
v_p^{VE} \qquad \text{if} \quad 0\leq \Delta \leq 1
\end{cases}
    \label{wall normal particle motion}
\end{equation}

Using equations \eqref{wall parallel particle motion}, \eqref{Us}, and \eqref{wallnormalvp} -- \eqref{wall normal particle motion} in the dynamical system \eqref{force free motion}, We have thus established a formalism for computing particle trajectories in the presence of channel wall effects for arbitrary Stokes background flow. \textcolor{black} {To the authors' knowledge, the present work is the first to formulate a closed hydrodynamics-based equation of motion for particles  entrained in wall-bounded Stokes flow.}

\section{Results and discussion}\label{sec: quantify results}

\subsection{Particle motion in Moffatt Eddy flow}\label{limit cyles}

We now use this formalism to discuss the fate of a neutrally buoyant spherical particle placed in
a vortical Moffatt flow. We only briefly mention that the computations confirm that the symmetric flow field 
\textcolor{black}{\eqref{streamfunction symmetric moffatt}} induces closed trajectories for any initial condition (see Fig.~\ref{Fig: Symmetric Moffatt}b) because of the equal and opposite effects of both walls. As our focus lies on the permanent displacement of particles, we concentrate in the following on the antisymmetric flow
given by (the real part of) the stream function $\psi$ of \eqref{streamfunction antisymmetric moffatt}, with
$u(x,y)=\mathcal{R}(\frac{\partial\psi}{\partial y})$, $v(x,y)=\mathcal{R}(-\frac{\partial\psi}{\partial x})$.

Without loss of generality, we will discuss trajectories of particles placed in the lower half of the channel, interacting more strongly with the lower wall at $y=-1$, though the influence of both walls is taken into account, see \eqref{wallnormalvp}. 

First, it is easy to see that if the flow field $(u,v)$ consists of closed (vortex) streamlines, the Faxen trajectories given by equation \eqref{Faxen velocity} must also close. Thus, the Faxen trajectory field can be interpreted as an altered 'incompressible flow field'. For small $a_p$, this altered flow $\boldsymbol{u}_{p,F}$ is a perturbation of the Moffatt flow.

Any permanent particle displacement (non-closing trajectories) is thus a result of the wall correction $\boldsymbol{W}$, a further perturbation on the reference field $\boldsymbol{u}_{p,F}$. Note that while it is tempting to model only half of the channel and focus on, say, one of the lower half vortices in Fig.~\ref{Fig3}(a) bounded by a no-slip wall at $y=-1$ and a no-stress wall at $y=0$, the disturbance flow from the particle will violate the latter boundary condition. We also verify that for small enough particles, the results of this approach are indistinguishable from the formalism for two no-slip walls (see Supplementary Information).

\begin{figure}
    \centering
\includegraphics[height=3.9cm]{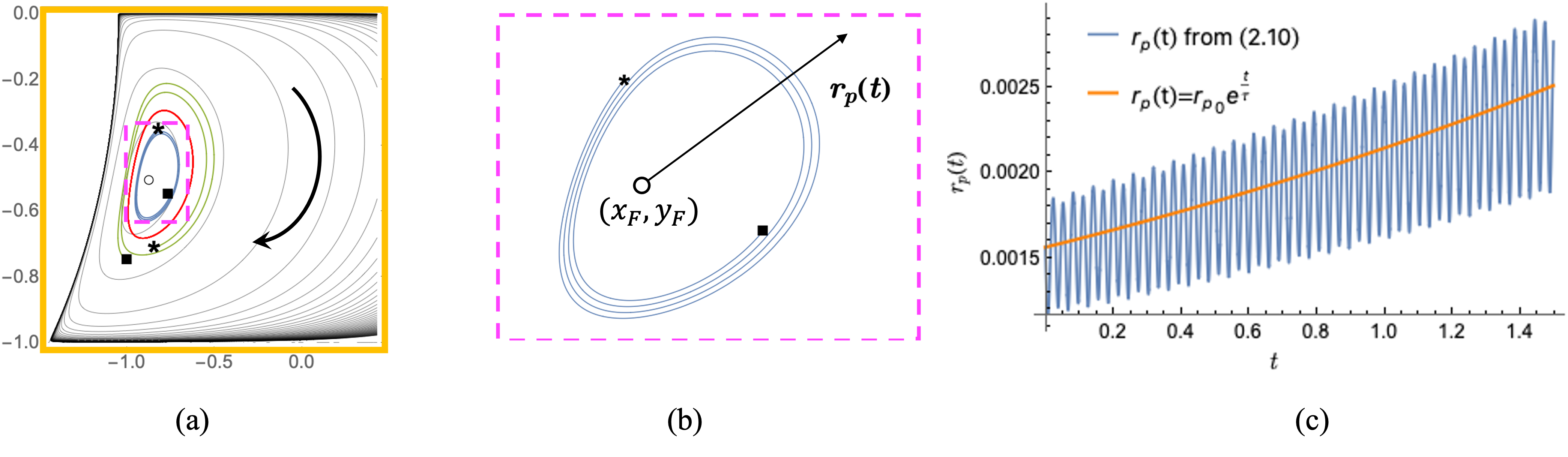}
    \caption{Particle trajectories near a stable limit cycle. (a) Particles ($a_p=0.2$) spiral out (blue) or spiral into (green) a stable limit cycle (red) in a clockwise eddy. $\circ$ indicates the unstable fixed point, $\blacksquare$ indicates the starting points of the particles, and $\boldsymbol{\ast}$ indicates the end points. (b) Close-up indicating radial distance of the particle $r_{p}(t)$ from the fixed point of the Faxen field. (c) Analytical results for the average of $r_{p}(t)$ match the solution of the dynamical system \eqref{wallnormalvp}; $\tau$ is defined in \eqref{spiral rate}.}
     \label{Fig4}
\end{figure}

In an antisymmetric Moffatt eddy, the vortex symmetry is broken in both the $x$ and $y$ directions so that there is no {\em a priori} reason for particles to follow closed trajectories. Let us first focus on initial conditions inside a clockwise vortex (yellow frame in Fig.~\ref{Fig3}(a), isolated in Fig.~\ref{Fig4}(a)). Solving \eqref{force free motion} for reasonably small particle size ($a_p\leq 0.2$), the following observations can be made: (i) particles initially placed near the vortex center follow trajectories that spiral away from the center (blue in Fig.~\ref{Fig4}(a); also see the close-up of Fig.~\ref{Fig4}(b)); (ii) particles initially placed near the outer edge of the vortex follow trajectories that spiral inwards (green in Fig.~\ref{Fig4}(a)); (iii) the spiraling is significantly slower for smaller $a_p$.

This suggests the presence of an unstable fixed point near the vortex center (open circle in Fig.~\ref{Fig4}(a)) and the existence of a stable limit cycle at a finite distance from the wall (red in Fig.~\ref{Fig4}(a)). As particles complete cycles in the vortex, the wall encounters have a cumulative effect that pushes them toward one well-defined closed trajectory, suggesting the possibility of systematic particle manipulation and accumulation even for force-free spheres in zero-$Re$ Stokes flow. In what follows, we shall quantify these effects.

\subsection{Particle motion near a Faxen field fixed point} \label{subsec: near Faxen field fixed point}
\subsubsection{Linear stability analysis}

The observed (slow) spiraling away from the fixed point must result from the wall corrections added to the Faxen field, whose trajectories are closed. Linearizing around the fixed point of the Faxen field yields quantitative predictions of the spiraling rate. Note that for a particle of any reasonable size $a_p\ll 1$ near the fixed point position (at $y\approx -0.5$), the gap measure will be $\Delta\gg 1$, so that wall effects are accurately described using the large-$\Delta$ asymptotics of \eqref{large ABCD}. The leading-order wall correction terms are then considerably simpler,
\begin{equation}
    W_{x,large}(x,y)= -\frac{c a_{p}^3}{(y+1)^3} u_{p,F}(x,y)\,,
    \label{large delta parallel correction}
\end{equation}
\begin{equation}
    W_{y,large}(x,y)= -\frac{15}{16}\frac{a_{p}^3}{(y+1)^2}\frac{\partial v_{p,F}(x,y)}{\partial y}\,,
     \label{large delta normal correction}
\end{equation}
where the constant $c$ can vary depending on the specific flow, but is ${\cal O}(1)$ \citep{goldman1967slow2, ghalia2016sphere}, cf.\ Sec.~\ref{subsection: wall parallel particle}.
Near the fixed point of the Faxen field, the $\frac{\mathcal{B}_{large}}{\mathcal{A}_{large}}$ term in $W_y$ dominates the others ($\frac{\mathcal{B}_{large}}{\mathcal{A}_{large}} = \mathcal{O}(a_p^3),\frac{\mathcal{C}_{large}}{\mathcal{A}_{large}} = \mathcal{O}(a_p^5),\frac{\mathcal{D}_{large}}{\mathcal{A}_{large}} = \mathcal{O}(a_p^4)$) and is the only one contributing to $\mathcal{O}(a_p^3)$ in \eqref{large delta normal correction}. This term is still of higher $a_p$ order than the Faxen correction so that, in the limit of large $\Delta$ and small $a_p$, the wall correction is perturbatively small.

We thus linearize \eqref{force free motion} around the Faxen field fixed point ($x_F$,$y_F$) (given by $\boldsymbol{v}_{p,F}=0$) 
and obtain
\begin{gather}
 \begin{bmatrix} \dot x_{p} \\ \dot y_{p} \end{bmatrix}
 =\nabla{{\boldsymbol{u}_{p}\big|_{(x_F,y_F)}}}
  \begin{bmatrix}
  x-x_{F} \\
   y-y_{F}
   \end{bmatrix}
\end{gather}
The matrix $\nabla{\boldsymbol{u}_{p}}\equiv A_1$ of this dynamical system can be decomposed as
\begin{equation}
    \textbf{$A_1$}=\textbf{$A_F$}+\textbf{$S$}\,,
\end{equation}
where \textbf{$A_F$} is the Jacobian of the Faxen field:
$$
\textbf{$A_F$}=\begin{bmatrix}
  \frac{\partial u_{p,F}(x,y)}{\partial x}&\frac{\partial u_{p,F}(x,y)}{\partial y} \\
    \frac{\partial v_{p,F}(x,y)}{\partial x} &\frac{\partial v_{p,F}(x,y)}{\partial y}
\end{bmatrix}\Bigg|_{(x_F,y_F)}$$
and \textbf{$S$} is due to the wall corrections,

\[\textbf{$S$}=\begin{bmatrix}
  \frac{\partial W_x(x,y))}{\partial x}&\frac{\partial W_x(x,y))}{\partial y} \\
    \frac{\partial W_y(x,y))}{\partial x}&\frac{\partial W_y(x,y))}{\partial y}
\end{bmatrix}\Bigg|_{(x_F,y_F)}\]

The eigenvalues of $A_1$ are
\begin{equation}
\lambda_{1,2}^{A_1}=1/\tau\pm i\omega_1\,,
\label{eigenvalue}
\end{equation}
where the real part $1/\tau$ is due to the wall correction $S$ only, as the fixed point of the incompressible Faxen field is a center.

The imaginary part $\omega_1$ is the angular frequency of the spiraling motion, which differs only perturbatively from that of the Faxen field, $\omega_{1}=\omega_{F}+\mathcal{O}(a_p^{3})$, where $\omega_{F}\equiv \sqrt{det(A_F)}$. To leading order, the frequency can be evaluated directly from the background flow, i.e., 
\begin{equation}
\omega_{F}=\omega_{0}+\mathcal{O}(a_p^{2})\equiv \sqrt{\frac{\partial u_(x,y)}{\partial x}\frac{\partial v(x,y)}{\partial y}-\frac{\partial u(x,y)}{\partial y}\frac{\partial v_(x,y)}{\partial x}}+\mathcal{O}(a_p^{2})\,.
\label{omega0}
\end{equation}

\subsubsection{Analytical prediction of particle spiraling rate} \label{subsection: spiral rate}
The real part of the eigenvalue \eqref{eigenvalue} translates into an exponential growth rate of the radial distance $r_p$ of the particle from $(x_F,y_F)$, i.e., $1/\tau = Tr(A_1)/2= Tr(S)/2$, i.e., 
\begin{equation}
    \frac{1}{\tau}=\frac{1}{2}\left(\frac{\partial W_{x,large}(x,y)}{\partial x}+\frac{\partial W_{y,large}(x,y)}{\partial y}\right)\bigg|_{(x_F,y_F)}
\end{equation}

Using the simplified wall corrections \eqref{large delta parallel correction} and \eqref{large delta normal correction}, neglecting higher orders of $a_p$, and using incompressibility, we obtain an explicit expression for the characteristic radial growth rate in terms of the background flow field only,
\begin{equation}
    \frac{1}{\tau}=\frac{1}{r_{p}}\frac{dr_{p}}{dt}=\frac{a_{p}^3}{32(y+1)^3}\left((16c+30)\frac{\partial v(x,y)}{\partial y}-15(y+1)\frac{\partial ^2 v(x,y)} {\partial y^2}\right)\bigg|_{(x_F,y_F)}
    \label{spiral rate}
\end{equation}
 
The resulting particle motion $r_{p} (t) =r_{p0}e^{\frac{t}{\tau}}$ from \eqref{spiral rate} with $c=5/16$ is shown in Fig.~\ref{Fig3}(c), demonstrating excellent agreement with the average numerically determined distance from the fixed point of the Faxen field $(x_F,y_F)$ (oscillations are due to the non-circular shape of the orbit).

Although this spiraling rate will change quantitatively with $c$, any ${\cal O}(1)$ values of $c$ will give very similar values of $1/\tau$ (choosing $c$ a factor of 2 larger or smaller only changes the spiraling rate by $\pm 12\%$). In the following, we will use the linear-shear value $c=5/16$, as it is the physical choice for particles at smaller $\Delta$, which we will discuss below.

Near the fixed points of different vortices, the expression \eqref{spiral rate} is unchanged except for overall factors of the powers of $-\zeta$ (neighboring vortices having opposite orientation). Likewise, \eqref{omega0} remains valid up to powers of $\zeta$. Thus, a convenient dimensionless measure for the spiraling trajectories is $\beta\equiv |\omega_0\tau|$, valid for all vortices:
\begin{equation}
    \beta=|\omega_0\tau|\approx \frac{0.337}{a_{p}^3} 
    \label{radial spiraling rate}
\end{equation}
For $a_p\ll 1$, the value $\beta \gg 1$ represents the number of orbits around a Faxen field fixed point a particle travels until its radial distance changes significantly. 
If $a_p$ is 0.1, for instance, such characteristic particle displacement accumulates over about 300 cycles in the vortex.

\subsection{Particle motion and manipulation in a clockwise vortex}

\begin{figure}
    \centering
\includegraphics[height=12cm]{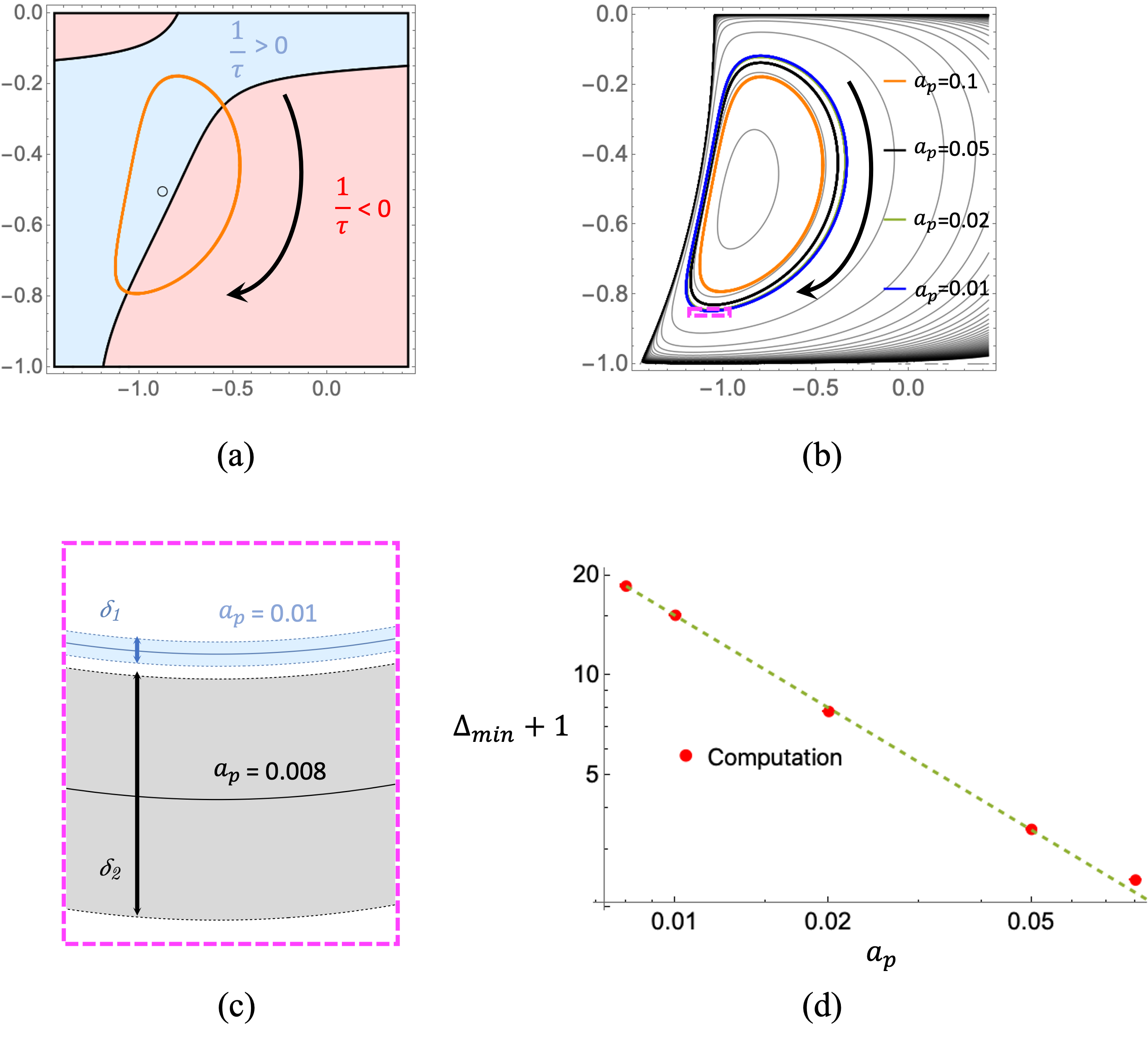}
    \caption{(a) Plot of the zero contour of $1/\tau$ together with the stable limit cycle of an $a_p$= 0.1 particle. $\circ$ indicates the unstable fixed point. (b) Particle size dependence of stable limit cycle location. (c) Close-up of limit cycles for $a_p=0.01, 0.008$ showing the bandwidths of uncertainty $\delta_{1}\approx 8\cdot 10^{-5}$, $\delta_{2}\approx 8\cdot 10^{-4}$. (d) The minimum gap between the particle and the wall obeys a power law in particle size: $\Delta_{min}+1\propto a_{p}^{\alpha}$, where $\alpha\approx -0.92$.}
\label{Fig5}
\end{figure}

As empirically shown by the red line in Fig.~\ref{Fig3}(a), the spiraling out of particles from the unstable fixed point eventually settles onto an asymptotically closed trajectory (a stable limit cycle), also reached from initial conditions closer to the wall, resulting in an inward spiral. 
As a motivation for the existence of this limit cycle, we compute the real part of the eigenvalues of the linearized dynamical system in the entire vortex region, i.e., \eqref{spiral rate} for arbitrary $(x,y)$. Figure~\ref{Fig5}(a) shows that a particle spiraling out from $(x_F,y_F)$ at first encounters only positive rates of radial growth, but then a greater and greater part of the trajectory is taken up by points with negative growth rate. Eventually, on the limit cycle, the integral effect of positive and negative growth balances.

How does the limit cycle location depend on particle size? As the spiraling rate decreases dramatically with smaller $a_p$ according to \eqref{radial spiraling rate}, direct forward integration of \eqref{force free motion} to the limit cycle is very time-consuming. Instead, we adopt a bisection scheme, integrating from an initial $x$-position until the same $x$ coordinate is reached again, registering the change $\Delta y$ in the $y$ coordinate. Iterating between initial conditions of positive and negative $\Delta y$, we find the position of periodic trajectories with great accuracy.

We plot the stable limit cycle locations for different particle sizes in Figure \ref{Fig5}(b). Note that for reasonably small $a_p$ even the point of closest approach to the wall has $\Delta=\Delta_{min}\gg 1$, so that using the large-$\Delta$ approximations of $\mathcal{A},\mathcal{B},\mathcal{C},\mathcal{D}$ from Appendix~\ref{appendix B} is quantitatively accurate.
As the particle gets smaller, the limit cycle grows, with a minimum distance $h_{min}$ closer to the wall, though $h_{min}$ decreases very slowly for very small $a_p$. Due to the exponential $x$-dependence of the flow field components resulting from \eqref{streamfunction antisymmetric moffatt}, we need to control for numerical errors in forward integration. Carefully evaluating (for a given $a_p$) the limit cycles starting from different initial positions, we obtain a band of uncertainty around the mean limit cycle. Figure~\ref{Fig5}(c) shows that this uncertainty $\delta$ increases as $a_p$ decreases. With the standard numerical accuracy and scheme we used, uncertainty bands begin to overlap for $a_p\lesssim 0.005$. Accurate data for smaller $a_p$ could be accessed with more powerful algorithms or CPUs, but this is not our focus here. Restricting ourselves to $a_p\geq 0.008$, we show $h_{min}/a_p=\Delta_{min}+1$ as a function of $a_p$ in Fig.~\ref{Fig5}(d), demonstrating an accurate power law (all uncertainties are below the symbol size) of the form
\begin{equation}
(\Delta_{min} + 1) \propto a_p^\alpha\,,
    \label{deltaminpower}
\end{equation}
with an exponent $\alpha\approx -0.92$. Note that $\Delta_{min}$ diverges as $a_p\to 0$ so that the $\Delta\gg 1$ limit becomes ever more accurate. Indeed, none of the quantitative results of Fig.~\ref{Fig5}(d) change when the full or asymptotic expressions for the wall effects are used. The scaling implies $h_{min}\propto a_p^\eta$ with $\eta\approx 0.08$, confirming the very slow approach of the limit cycles towards the wall. 

For particle sizes relevant to microfluidics, this effect means that there is a practical boundary for how close to the wall the particles can approach. Taking the characteristic (half-width) length of the channel to be 50$\mu$m, a $1\mu$m particle ($a_p=0.02$) will not approach the wall any closer than 7.8$\mu$m. Moreover, using the locations of stable limit cycles to separate particles becomes very difficult for small particles. For practical situations, the difference between the $h_{min}$ for a 1 $\mu m$ particle and a 2.5 $\mu m$ particle is only 0.7$\mu m$. \textcolor{black}{Separation by size can be further compromised by the presence of Brownian motion. Using characteristic time scales $\tau$ from \eqref{radial spiraling rate} to estimate positional uncertainty due to Brownian diffusion, we find that an $a_p=5\mu$m particle is hardly affected, while the position of a strongly colloidal $a_p=1\mu$m particle will be spread out over several $\mu$m.}
  
\textcolor{black}{Despite these caveats}, the ability to concentrate particles on a limit cycle trajectory without inertial effects purely because of the background flow geometry is of fundamental interest. 
It is encouraging that the location of this limit cycle for small particles can be determined entirely within the large $\Delta$ approximation, i.e., without the intricate details of near-wall corrections or lubrication limits. This gives confidence in not only the qualitative but also the quantitative description of the phenomenon: when placed in certain bounded vortical Stokes flows, small spherical particles, even when neutrally buoyant, will eventually accumulate on well-defined closed trajectories. 

\begin{figure}
    \centering
\includegraphics[height=4.2cm]{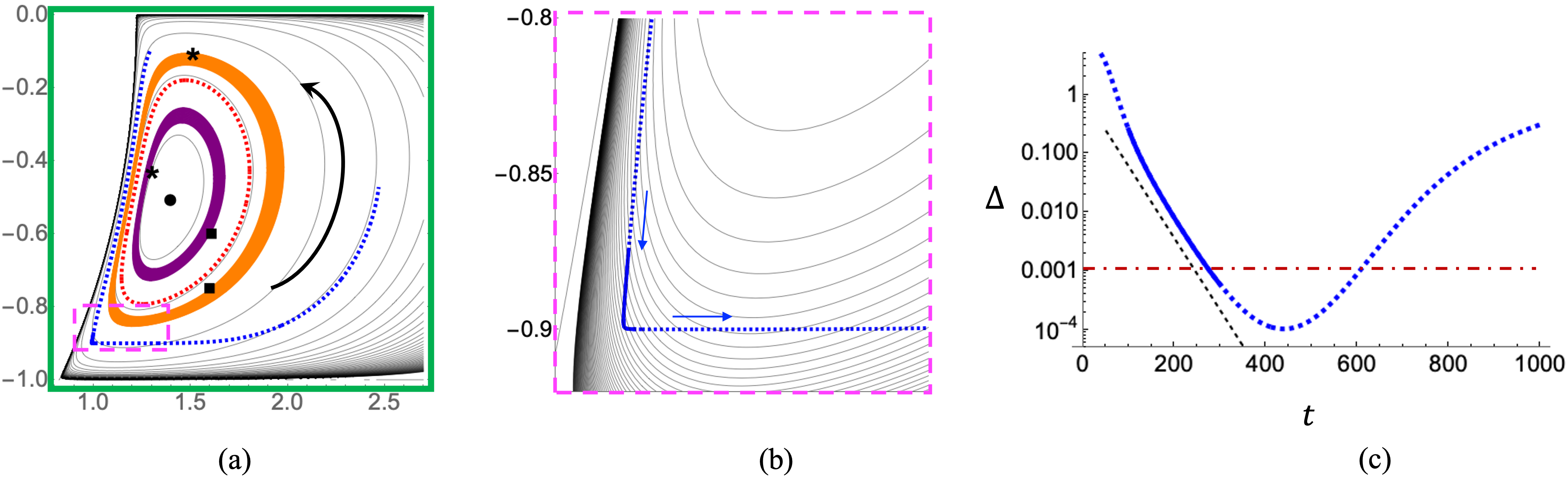}
    \caption{Particle trajectories near an unstable limit cycle. (a) Particles ($a_p=0.1$) spiral out (orange) towards the wall or spiral into (purple) a fixed point from an unstable limit cycle (dashed red) in a counterclockwise eddy. A particular spiraling-out trajectory is shown in blue. $\bullet$ indicates the stable fixed point, $\blacksquare$ indicates the particle starting points, and $\boldsymbol{\ast}$ indicates the particle end points. (b) Close-up of the close approach to the wall of the trajectory from (a). The solid portion of the trajectory shows approximately exponential thinning of the gap, shown in the semilog plot of (c). The black dashed line indicates the exponential behavior from the wall-expansion approximation \eqref{logdelta}. The dot-dashed red line corresponds to a surface-to-surface approach of 5\,nm distance for a 5 $\mu$m particle in a channel of 50 $\mu$m half-width.}
     \label{Fig6}
\end{figure}

\subsection{Particle motion and manipulation in a counterclockwise vortex}\label{subsec: particle sticking}

Any vortex adjacent to a clockwise vortex like the one discussed in Sec.~\ref{subsec: near Faxen field fixed point} is counterclockwise and congruent in geometry. For example, the vortex indicated by the green frame in \textcolor{black}{Fig.~\ref{Fig3}(a)} has flow exactly reversed from the yellow-framed vortex (and a factor $\zeta$ slower). Because of the time reversibility of Stokes flow and the fact that all wall effects result from the background Stokes flow and its derivatives, the behavior of particles on trajectories is also time-reversed. Thus, the fixed point in a counterclockwise vortex is stable, 
$\frac{1}{\tau}$ changes sign, and $\beta$ stays the same by definition. The limit cycle for a given $a_p$ is unstable but is congruent in shape with the stable cycle discussed before. Particles spiral inward from the unstable limit cycle towards the fixed points but spiral outward when placed outside the limit cycle (Figure \ref{Fig6}(a)). The latter case is of prime interest because it allows particles to approach the wall more and more closely. As the gap between particle and wall diminishes, any short-ranged intermolecular attractive force (e.g.\ Van der Waals force) whose reach is often in the nm range \citep{hirschfelder1954molecular, batsanov2001van} can take over and lead to attachment (sticking) of the particle to the wall. This general mechanism (hydrodynamics allows a particle to get close enough to a wall to stick by short-ranged attraction) has been acknowledged before \citep{friedlander2000smoke, humphries2009filter}, but the case discussed here allows for quantitative predictions.

Figure~\ref{Fig6}(b) exemplifies a particle trajectory that approaches very close to the wall, with a nearly normal initial approach and \textcolor{black}{a subsequent piece of trajectory nearly parallel to the wall}. \textcolor{black}{Here, $\Delta\ll 1$ characterizes the trajectory and this is the only scenario in the present work where the variable expansion method for particle normal velocity \eqref{wyVE} is necessary.} The semi-logarithmic plot vs.\ time in Fig.~\ref{Fig6}(c) shows that the "near-parallel" portion contains a stretch of approximately exponential decay of the (relative) gap $\Delta$. Indeed, this behavior follows from a series to leading order in small $\Delta$ of \eqref{wyVE}, i.e., the wall-expansion limit of $y_E\to -1$. Here, \eqref{wallnormalvpmodify} simplifies to
\begin{equation}
    \frac{d \Delta}{dt}=1.6147a_p\kappa(x)\Delta\,,
    \label{logdelta}
\end{equation}
where $\kappa(x) = \frac{\partial^2v}{\partial y^2}(x,-1)$ is the background flow curvature at the wall, and the prefactor was first derived in \citet{goren1971hydrodynamic} and confirmed in \citet{rallabandi2017hydrodynamic}. The trajectory part showing the exponential approach (the solid portion in Fig.~\ref{Fig6}(b,c)) indicates slow motion in the $x$-direction, so that $\kappa(x)$ is nearly constant. We fix the $x$-value as that of the turning point (point of maximum curvature) of the trajectory, here $x\approx 0.991$. Using this value in \eqref{logdelta} gives an exponential behavior (dashed line in Fig.~\ref{Fig6}(c)) in good agreement with the approach of the trajectory to the wall. This behavior remains robust to changes of the exact modeling of the expansion point locations $y_E$ in \eqref{yeofy}.

This exponential approach to a wall accords with the general derivations of Brady et al.\ showing that the gap between solid surfaces in Stokes flow can never vanish in finite time  \citep{brady1988stokesian,claeys1989lubrication,claeys1993suspensions}. 
In a practical situation, the rapidly decaying gap width leads to sticking by short-range interaction at a well-defined position after a well-defined time. If Fig.~\ref{Fig6}(c) pertains to a microfluidic situation exemplified by a channel half-width of 50$\mu$m and a particle radius of 5$\mu$m, the red dashed line indicates a gap of 5\,nm between particle and wall, at the low end of the typical range of van der Waals forces \citep{israelachvili1974nature}. The particle can, therefore, be expected to stick at a time and location entirely determined by the geometry of the background flow. Thus, we demonstrate here how, in a pure Stokes flow, particles can be driven toward boundaries and forced to stick to a wall in predictable locations. Note that conceptually, a counterclockwise vortex allows for the concentration of randomly distributed particles in two locations: the stable fixed point near the center of the vortex, and the well-defined small region of wall sticking. Both alternatives provide practically relevant protocols for filtering in microfluidic devices. \textcolor{black}{Particles placed inside the unstable limit cycle will be concentrated at the fixed point, while  particles placed outside the unstable limit cycle will be deposited onto a surface.}

\textcolor{black}{Our single-particle formalism implictily assumes the limit of dilute particle concentration in a microfluidics application. Including effects of particle-particle interaction is a desirable future extension of this approach in order to assess non-inertial effects in particle-laden flows important to practical applications \citep{guha2008transport}. }

\section{Conclusions}\label{sec: conclusion}

We have quantitatively shown that force-free particles placed in an internal Stokes flow can be systematically displaced in a variety of ways through purely hydrodynamic interactions with the enclosing walls. Since a particle cannot contact a wall in finite time, a typical trajectory has a portion of approaching the wall and a portion of receding, governed by the background flow and the wall corrections derived from it. If the approaching and receding parts of the flow are symmetric, displacement effects off a streamline will cancel out. Thus, a Stokes-flow-inducing net particle displacement must break wall-parallel symmetry.

For practical use, vortical flows are advantageous, where the effects of many approaching/receding events can accumulate. However, if the vortex is symmetrically confined between walls, the net displacement effects induced by both walls will again cancel, and the particle trajectory must close. For net cumulative displacement, the vortex must, therefore, break wall-normal symmetry as well, for example by being confined between no-slip and no-stress boundaries. A net-transport internal Stokes flow thus breaks symmetries in both directions by particle-wall interactions, which act as perturbations on the Faxen flow trajectories. This is in agreement with recent experimental results that record permanent displacements of fibers passing by obstacles of symmetry-broken shape \citep{li2024dynamics}. For such non-spherical particles (whether rigid or elastic), the effect of a single particle-boundary encounter could be much enhanced, as the wall effects affect different parts of the particle differently, and the overall torque balance will influence the resulting displacement. The study of this case will be the subject of future work.

Our formalism predicts how the particles approach or recede from fixed points at the center of the vortices, and how the particles accumulate at stable limit cycles, whose locations are particle size-dependent. When placed in certain bounded vortical Stokes flows, small spherical particles, even when neutrally buoyant, will eventually accumulate on well-defined closed trajectories. In practical applications, the accumulated particles can be induced to aggregate or react with each other. We can also predict how particles move away from unstable limit cycles towards boundaries. Generically, such trajectories must eventually lead to exponential thinning of the fluid layer between particle and wall. Thus, sub-micron distances are reached, leading to sticking in predictable locations by short-range forces. Such adhesion to walls in Stokes flows adds a simple and controllable tool to studies of sticking phenomena, which has been studied for some time. Researchers have investigated particles captured, filtered, and deposited on surfaces in low Reynolds flow by short-range interactions with the wall when approaching \citep{arias2019low,espinosa2012particle}. Applications include fiber filtration \citep{tien1977chainlike, myojo1984experimental} and fog meshes \citep{park2013optimal} inspired by moisture-collecting desert beetles \citep{parker2001water, king2022beetle}. This is also related to the functionality of facemasks in the recent COVID-19 pandemic \citep{howard2021evidence}. However, most of the work emphasizes the dependence of particle deposition probabilities on the finite Stokes number in the flow, while few studies focus on the Stokes flow regime or acknowledge that any filtration effect is possible in this limit.

Let us estimate the rate of particle drift across streamlines in dimensional terms. Taking the length scale (channel half-width) as $H=50\mu$m, we want to ensure that the channel Reynolds number $Re=HU/\nu$ for aqueous solutions ($\nu\sim 10^{-6}$m$^2$/s) is less than $10^{-2}$ along the particle trajectories, even in regions of the highest speed $U$. Given this constraint and assuming $a_p=5\mu$m, the time scale $\tau$ from \eqref{spiral rate} is $\approx 57$s so that a significant radial displacement is expected on a time scale of minutes. For larger particles, this time scale decreases rapidly ($\propto a_p^{-3}$), and the process can also be sped up by using a solution of higher viscosity. 

The Moffatt eddy flow discussed here is convenient because of its closed analytical form. However, it would be extremely difficult to set up experimentally due to the exponential decay of the flow speed with $x$ and the need to drive the flow far away from the field of view. The exponential dependence of speed on coordinates also makes the Reynolds number constraint for Stokes flow extremely stringent. Nevertheless, the principles of displacement and the scaling of the wall effects are not specific to this flow and will be robust in any symmetry-breaking vortical Stokes flow, \textcolor{black}{and the modeling equations for the crucial wall-normal displacement such as \eqref{wyfull} and \eqref{wyVE} are valid for arbitrary background flow.
}

Cavity flows are a class of vortical flows that is practically used in microfluidic devices at low and vanishing Reynolds numbers. In the Stokes limit, they can be expressed as infinite sums of Moffatt solutions \citep{shankar1993eddy, shankar2000fluid} and can be driven by rotating cylinders \citep{hellou1992cellular, hellou2001sensitivity} or by superimposed transport flows \citep{hellou2011stokes}. Because of the linearity of the Stokes equations and the velocity corrections, such flows do not pose principal difficulties to the present formalism. 
These finite-domain modifications of Moffatt's solutions have a less severe ratio of driving velocity to the eddy velocity scale (the analog of $e^{p\pi/q}$), and it is thus much easier to fulfill the conditions of small $Re$.

It is instructive to compare the scaling of the migration velocity of small particles far from the wall ($\Delta\gg 1$) with that of approaches using particle inertia. \textcolor{black}{In a number of approaches quantifying the effect of steady inertia \citep{ho1974inertial,schonberg1989inertial,asmolov1999inertial,di2007continuous,hood2015inertial} the migration speed for force-free particles is proportional to $a_p^3$, the same scaling as} our result \eqref{large delta normal correction}. In approaches using oscillatory inertia \citep{agarwal2021unrecognized,agarwal2024density}, the scaling varies from $a_p^3$ for small Stokes number to $a_p^4$ for large Stokes number. Thus, the strictly non-inertial effects described in the present work show scaling equal to or even more favorable than inertial techniques for small particles.   
\textcolor{black}{It should be noted that at low (non-zero) $Re$ the walls of a microfluidic device will almost always be close enough to the particle to enable modeling as in \citet{ho1974inertial} or \citet{hood2015inertial}, i.e., the effects of inertia at large distances are not present.  This is why Saffman lift \citep{saffman1965lift}, an unbounded-flow effect, should not be directly compared with these results.} 

In closing, we note that oscillatory-flow microfluidic setups used for fast particle manipulation often result in both inertial forces on particles \citep{agarwal2021unrecognized,agarwal2024density}
and in the generation of steady vortical streaming flow of low Reynolds number \citep{rallabandi2014two,thameem2017fast,rallabandi2017hydrodynamic}. The effects discussed in the present work can thus be exploited and optimized together with inertial effects, leading to better protocols for the accumulation, concentration, deflection, and sorting of microparticles.

\section*{Acknowledgments}
\begin{acknowledgments}
The authors acknowledge valuable and inspiring conversations with John Brady, Camille Duprat, Anke Lindner, Bhargav Rallabandi, and Howard Stone.\\ Declaration of Interests: The authors report no conflict of interest.
\end{acknowledgments}

\section*{Supplementary information}

\appendix

\section{}\label{appendix A}
To get the shear coefficient of the particle parallel velocity $f(\Delta)$, we start with the form from \citet{pasol2011motion}: 
\begin{equation}
    \begin{aligned}
    f(\Delta)&=1-\left[1-a\log(1-\frac{1}{1+\Delta})-b_1(\frac{1}{1+\Delta})-b_2(\frac{1}{1+\Delta})^2 \right. \\& \left. -b_3(\frac{1}{1+\Delta})^3-b_4(\frac{1}{1+\Delta})^4\right]^{-1}
    \label{pasoldelta}
    \end{aligned}
\end{equation}

We take a series expansion of $f(\Delta)$ as $\Delta\to\infty$ to the order $\Delta^3$:
\begin{equation}
 \begin{aligned}
&f(\Delta)=-\frac{-a+b_1}{\Delta}-\frac{a^2+a(\frac{1}{2}-2b_1)-b_1+b_1^{2}+b_2}{\Delta^2}\\&-\frac{-a^3+b_1-2b_1^2+b_1^3+a^2(-1+3b_1)-2b_2+2b_1b_2-\frac{1}{3}a(1-9b_1+9b_1^2+6b_2)+b_3}{\Delta^3}\\&+H.O.T.
 \end{aligned}
\end{equation}
According to \citet{goldman1967slow2} as $\Delta\to\infty$, $f(\Delta)\simeq 1-\frac{5}{16}\Delta^{-3}$, which means that the order $\Delta^{-1}$ and $\Delta^{-2}$ must vanish, resulting in 
\begin{equation}
a-b_1=0\, \qquad a^2+a(\frac{1}{2}-2b_1)-b_1+b_1^{2}+b_2=0\,,
\end{equation}
so that $b_1=a$ and $b_2=\frac{a}{2}$.
With $b_1$ and $b_2$ substituted, 
matching with 
Goldman's large $\Delta$ asymptotic expression obtains
\begin{equation}
b_3=\frac{5}{16}+\frac{a}{3}\,.
\label{b3}
\end{equation}
We then take a series expansion of $f(\Delta)$ as $\Delta\to 0$ to leading order:
\begin{equation}
f(\Delta)\approx 1+\frac{2}{3a+2(-1+b_3+b_4)+2a\log(\Delta)}
\end{equation}
According to \citet{williams1994particle} as $\Delta\to0$, $f(\Delta)\simeq 1-\frac{1}{0.66-0.269\log(\frac{\Delta}{1+\Delta})}$. By matching all the parameters, we obtain $a=0.269$ and
\begin{equation}
\frac{1}{2}\left(-3a-2(-1+b_3+b_4 \right)=0.66
\label{b4}
\end{equation}
Combined with equations \ref{b3} and \ref{b4}, this determines
$b_3=-0.223$ and $b_4=0.159$. All parameters of \eqref{pasoldelta} are now specified, and the result is 
Eq.~\eqref{Us}.

\section{}\label{appendix B}

The scalar quantities $\mathcal{A}$, $\mathcal{B}$, $\mathcal{C}$ and $\mathcal{D}$ are dimensionless hydrodynamic resistances depending on $\Delta$. The analytical expressions for $\mathcal{A}$, $\mathcal{B}$, $\mathcal{C}$ and $\mathcal{D}$ are given (as infinite sums) in \citet{rallabandi2017hydrodynamic}. For large separations($\Delta \gg 1$), one obtains four hydrodynamic resistances at leading order:
\begin{equation}
    \mathcal{A}_{large} = 1+\frac{9}{8}\Delta^{-1}\,,\quad \mathcal{B}_{large} = \frac{15}{16}\Delta^{-1}\,,\quad \mathcal{C}_{large} = \frac{21}{32}\Delta^{-3} \,, \quad
    \mathcal{D}_{large} = \frac{1}{3}+\frac{3}{8}\Delta^{-1}
    \label{large ABCD}
\end{equation}
The ratios used in the equations for velocity corrections are, to leading order,
\begin{equation}
\frac{\mathcal{B}_{large}}{\mathcal{A}_{large}}\approx \frac{15}{16}\Delta^{-2}\,,\quad 
\frac{\mathcal{C}_{large}}{\mathcal{A}_{large}}\approx \frac{21}{32}\Delta^{-3}\,, \quad
\frac{\mathcal{D}_{large}}{\mathcal{A}_{large}}\approx \frac{1}{3}\,.
    \label{large delta ratios}
\end{equation}
For small separations($\Delta \ll 1$), one obtains four hydrodynamic resistances to leading order:
\begin{equation}
 \begin{aligned}
    \mathcal{A}_{small} &= \Delta^{-1}+\frac{1}{5}\log\Delta^{-1}+0.9713\,,\quad \mathcal{B}_{small} = \Delta^{-1}-\frac{4}{5}\log\Delta^{-1}+0.3070\,,\quad\\
    \mathcal{C}_{small} &= \Delta^{-1}-\frac{14}{5}\log\Delta^{-1}+3.7929 \,, \quad
    \mathcal{D}_{small} = \log\Delta^{-1}-0.9208\,.
    \label{small ABCD}
  \end{aligned}
\end{equation}
These results can be used to obtain the numerical prefactor of the wall-expansion limit equation \eqref{logdelta}.

\bibliographystyle{jfm}
\bibliography{jfm-instructions}

\end{document}